\documentclass[prl,aps,twocolumn]{revtex4}

\usepackage{amsmath}
\usepackage{bm}
\usepackage{graphicx}

\begin{document}

\title{Bloch Structures in a Rotating Bose-Einstein Condensate}

\author{Hiroki Saito}
\author{Masahito Ueda}
\affiliation{Department of Physics, Tokyo Institute of Technology,
Tokyo 152-8551, Japan \\
and CREST, Japan Science and Technology Corporation (JST), Saitama
332-0012, Japan
}

\date{\today}

\begin{abstract}
A rotating Bose-Einstein condensate is shown to exhibit a Bloch band
structure without the need of periodic potential.
Vortices enter the condensate by a mechanism similar to the Bragg
reflection, if the frequency of a rotating drive or the strength of
interaction is adiabatically changed.
A localized state analogous to a gap soliton in a periodic system is
predicted near the edge of the Brillouin zone.
\end{abstract}

\pacs{03.75.Kk, 05.30.Jp, 32.80.Pj, 67.40.Vs}

\maketitle

The Bloch band is crucial to our understanding of the behaviors of
periodic systems, such as electronic states in solids~\cite{Ziman}, atom
dynamics in optical lattices~\cite{Jessen}, and Cooper-pair tunneling in
small Josephson junctions~\cite{Likharev}.
The underlying physics common to all of these is the Bragg reflection that
occurs at the edges of the Brillouin zone.
The resultant Bloch band leads to various interesting phenomena, such as
Bloch oscillations~\cite{Ziman} (a particle in a periodic potential driven
by a weak constant force cannot be accelerated indefinitely but oscillates
in real space) and formation of gap solitons~\cite{Meystre} (localized
wave packets arising from the balance between negative-mass dispersion and
repulsive interaction); both phenomena have recently been observed with a
Bose-Einstein condensate (BEC) in an optical
lattice~\cite{Morsch,Eiermann}.

In this Letter, we show that yet another system --- a BEC confined in a
rotating harmonic-plus-quartic potential --- exhibits a Bloch band
structure, and investigate the associated novel phenomena.
Seemingly this system has no periodic structure, but may be considered to
be a quasi-1D periodic system in the following sense.
When the rotating frequency of the potential is high, the quartic
potential $\propto r^4$, together with the centrifugal one $\propto -r^2$,
produces a Mexican-hat shaped potential~\cite{Kasamatsu} whose minima form
a quasi-1D toroidal geometry.
Under such circumstances any perturbation $V(\theta)$ that is needed to 
drive the system into rotation by breaking the axisymmetry of the
potential serves as a ``periodic'' potential since $V(\theta) = V(\theta +
2\pi)$, where $\theta$ denotes the azimuthal angle.

In the present system, we will show that the Bragg reflection causes
vortex nucleation.
Since this process occurs adiabatically, in contrast to a method invoking
dynamical instabilities~\cite{Sinha,Madison01}, we can obtain the vortex
states without heating the atomic cloud.
A stirring technique to produce vortices without heating has also been
proposed in Refs.~\cite{Damski,Caradoc}.
Another scheme that adiabatically nucleates vortices is based on phase
engineering techniques~\cite{Dum,Williams,Nakahara}.
We will also show that a localized state is generated even with repulsive
interactions, which is analogous to gap solitons in periodic
systems~\cite{Meystre}.

We begin by discussing a BEC in a 1D ring to understand the essence of the
phenomena.
We assume that a rotating potential takes the form $V(\theta, t) =
\varepsilon \cos n (\theta - \Omega t)$.
In a frame rotating at frequency $\Omega$, the potential $V$ becomes
time-independent, and the Gross-Pitaevskii (GP) equation is given by
\begin{equation} \label{GP1D}
\left( -\frac{\partial^2}{\partial \theta^2} + i \Omega \frac{\partial
}{\partial \theta} + \varepsilon \cos n \theta  + \gamma |\psi_{\rm 1D}|^2
\right) \psi_{\rm 1D} = \mu \psi_{\rm 1D},
\end{equation}
where energy and time are measured in units of $\hbar^2 / (2 m R^2)$ and
$2 m R^2 / \hbar$ with $m$ and $R$ being the atomic mass and the radius of
the ring.
The wave function is normalized as $\int_0^{2\pi} |\psi_{\rm 1D}|^2
d\theta = 1$, and $\gamma$ characterizes the strength of interaction.

\begin{figure}[tb]
\includegraphics[width=8.4cm]{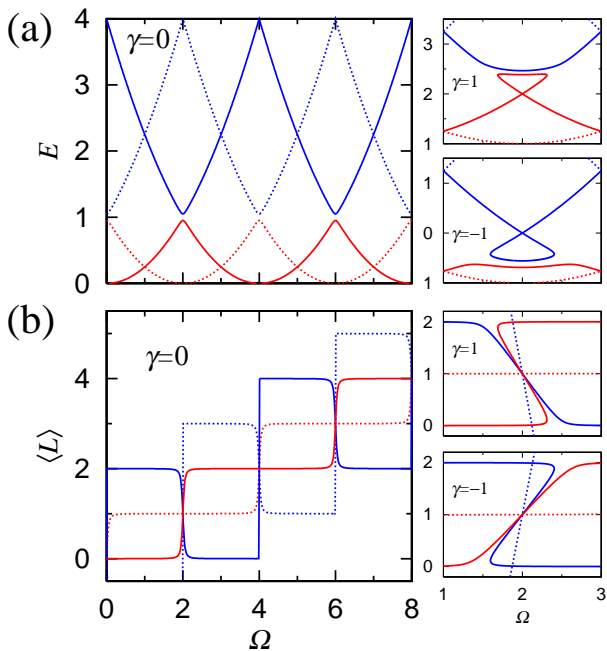}
\caption{
(a) $E \equiv \mu + \Omega^2 / 4$ and (b) $\langle L \rangle \equiv -i
\int \psi_{\rm 1D}^* \partial_\theta \psi_{\rm 1D} d\theta$ of the first
(red curves) and the second (blue curves) Bloch bands for the
noninteracting (main panels) and interacting (right panels) BECs in a 1D
ring described by Eq.~(\protect\ref{GP1D}) with $V(\theta) = 0.1 \cos 2
\theta$.
The solid and dotted curves refer to two independent branches.
}
\label{f:1D}
\end{figure}

First, let us consider the noninteracting case ($\gamma = 0$).
Introducing a new variable $\phi \equiv e^{-i \Omega \theta / 2} \psi_{\rm
1D}$, we transform Eq.~(\ref{GP1D}) into the Mathieu equation
$(-\partial_\theta^2 + \varepsilon \cos n \theta) \phi = E \phi$, where $E
\equiv \mu + \Omega^2 / 4$ plays the role of the ``total energy'' in our
Bloch band picture.
Then the solution satisfies $\phi(\theta + \Theta_0) = e^{i (p - \Omega /
2) \Theta_0} \phi(\theta)$, with $\Theta_0 \equiv 2\pi / n$ being the
period of the potential $V$, and $p$ an integer.
It follows that $p - \Omega / 2$ may be regarded as the quasimomentum,
indicating that we can move in the quasimomentum space by changing
$\Omega$.
Since the unit ``reciprocal lattice vector'' is $2\pi / \Theta_0 = n$, the
Bragg reflection occurs only between the states whose angular momenta
differ by $n$, $2n$, ... (Bragg's law).
Consequently there are $n$ independent branches corresponding to $p = 0,
\cdots, n - 1$ modulo $n$.
Figure~\ref{f:1D} shows $E$ and the angular momentum $\langle L \rangle =
-i \int_0^{2\pi} \psi_{\rm 1D}^* \partial_\theta \psi_{\rm 1D} d\theta$ as
functions of $\Omega$ for the first and second Bloch bands with $n = 2$
and $\varepsilon = 0.1$.
Two independent branches exist in this case, and are distinguished by the
solid and dotted curves.
The Bloch band structure in Fig.~\ref{f:1D} indicates that if we
adiabatically increase or decrease $\Omega$ across $\Omega = 2$, the
nonvortex state $\psi_{\rm 1D} \simeq 1 / \sqrt{2\pi}$ transforms to the
vortex state $\psi_{\rm 1D} \simeq e^{2 i \theta} / \sqrt{2\pi}$, and vice
versa.
The time scale of the adiabatic change must be much longer than the
inverse energy gap $\simeq \varepsilon^{-1}$.

When $|\gamma| \gtrsim \varepsilon$, the system exhibits hysteretic
behavior, and a loop structure emerges in the Bloch bands~\cite{Wu} as
shown in Fig.~\ref{f:1D}.
We see that the first (second) band forms a loop for $\gamma = 1 (-1)$;
then the adiabatic nucleation of vortices is not possible with increasing
(decreasing) $\Omega$ because the nonlinear Landau-Zener
transition~\cite{Wu} occurs no matter how slow the change of $\Omega$.
It can be shown that the states on the loop have dynamical instability
that breaks the symmetry of the system $|\psi_{\rm 1D}(\theta)| =
|\psi_{\rm 1D}(\theta + \Theta_0)|$.
This instability is similar to the one that forms gap solitons in a BEC in
an optical lattice~\cite{Konotop}.

We now consider a BEC in a harmonic-plus-quartic potential with a rotating
stirrer.
We assume that the axial trapping energy $\hbar \omega_z$ is much larger
than the other characteristic energies (a tight pancake-shaped trap) so
that the system is effectively 2D.
The external potential in the corotating frame of reference takes the form
$V({\bf r}) = r^2 / 2 + K r^4 / 4 + \varepsilon r^2 \cos 2 \theta$, where
$K$ is a constant and the third term $\varepsilon (x^2 - y^2)$ is a
stirring potential which can be produced by laser beams propagating in the
$z$ direction.
Such an anharmonic potential was theoretically considered in
Refs.~\cite{Kasamatsu,Lundh,Lundh04} and has recently been realized
experimentally~\cite{Bretin}.
We normalize the length, time, energy, and wave function by $d_0 \equiv
(\hbar / m \omega_\perp)$, $\omega_\perp^{-1}$, $\hbar \omega_\perp$, and
$\sqrt{N} / d_0$, respectively, with $\omega_\perp$ and $N$ being the 
frequency of the radial harmonic trap and the number of atoms.
The wave function is then normalized as $\int |\psi_{\rm 2D}|^2 d{\bf r} =
1$.
The time-dependent GP equation in the frame corotating with the stirring
potential at frequency $\Omega$ is given by
\begin{equation} \label{GP2D}
i \frac{\partial \psi_{\rm 2D}}{\partial t} = \left( -\frac{1}{2} \nabla^2
+ i \Omega \frac{\partial}{\partial \theta} + V({\bf r}) + g |\psi_{\rm
2D}|^2 \right) \psi_{\rm 2D},
\end{equation}
where $g = [\omega_z / (2\pi \omega_\perp)]^{1/2} 4\pi N a / d_0$
characterizes an effective strength of interaction in 2D~\cite{Castin}
with $a$ being the s-wave scattering length.

When $\Omega \gg 1$, the system described by Eq.~(\ref{GP2D}) can be
approximated by a quasi-1D ring~\cite{Kasamatsu}.
Assuming $\psi_{\rm 2D} \simeq f(r) e^{i \ell \theta} / \sqrt{2\pi}$, we
find that an effective potential for the radial wave function is given by
$\ell^2 / (2 r^2) + r^2 / 2 + K r^4 / 4$.
This potential has a minimum at $r \simeq (\ell^2 / K)^{1/6}$ for $\ell
\gg 1$, and the effective frequency around it is $\omega_{\rm eff} \simeq
\sqrt{6} (\ell K)^{1/3}$.
If $\omega_{\rm eff}$ is much larger than other characteristic
frequencies, the dynamics of $f(r)$ can be ignored.
After normalization of the time by $\rho^{-2} \equiv \int r |f|^2 r^{-2}
dr$, the equation of motion for $\theta$ reduces to Eq.~(\ref{GP1D}),
where $\gamma$ is given by $2 g \rho^2 \int r |f|^4 dr$.
Thus, the system described by Eq.~(\ref{GP2D}) exhibits the quasi-1D
circular flow for large angular momentum~\cite{Kasamatsu}, and we expect
that a Bloch band structure emerges.
We show below that this is true even for $\Omega \sim 1$ and $\ell \sim
1$.

\begin{figure}[tb]
\includegraphics[width=8.4cm]{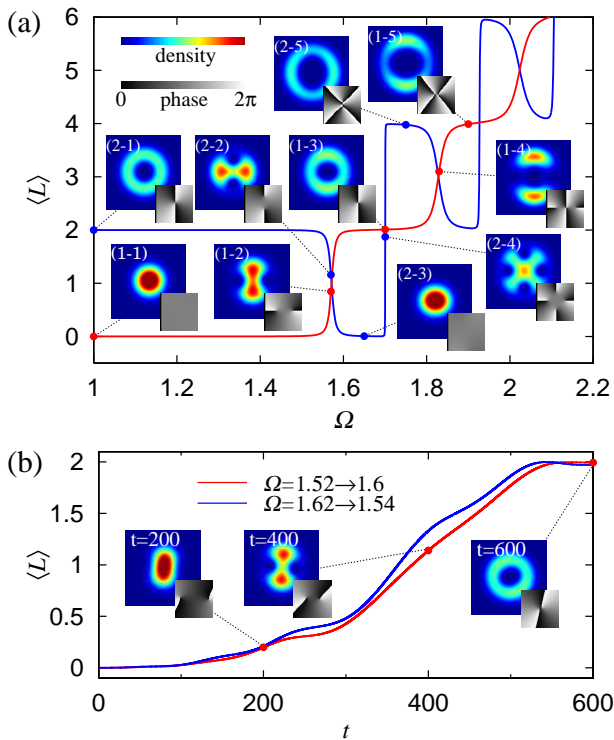}
\caption{
(a) Angular momentum $\langle L \rangle$ versus rotation frequency
$\Omega$ of the stationary states of the GP equation (\protect\ref{GP2D})
with $g = 0$, $K = 1$, and $\varepsilon = 0.02$.
The images (1-1)-(1-5) [(2-1)-(2-5)] correspond to the branch indicated by
the red (blue) curve.
(b) Time evolutions with $\lambda = 0.03$, where $\Omega$ is linearly
changed from 1.52 to 1.6 (red curve) and from 1.62 to 1.54 (blue curve)
during $0 \leq t \leq 600$.
The initial state is the ground state, and $\varepsilon$ is linearly
ramped up from 0 to 0.04 during $0 \leq t \leq 200$, kept at 0.04 during
$200 \leq t \leq 400$, and ramped down from $0.04$ to $0$ during $400 \leq
t \leq 600$ to suppress nonadiabatic disturbances.
The size of the images is $5 \times 5$ in units of $(\hbar / m
\omega_\perp)^{1/2}$.
}
\label{f:2Dnoint}
\end{figure}

Figure~\ref{f:2Dnoint} (a) shows the angular momentum $\langle L \rangle =
-i \int \psi_{\rm 2D}^* \partial_\theta \psi_{\rm 2D} d{\bf r}$ and the
density and phase profiles (insets) of the noninteracting stationary
states of Eq.~(\ref{GP2D})~\cite{note}.
We find that the behavior of $\langle L \rangle$ is similar to that in
Fig.~\ref{f:1D} (b).
When we start from the ground state with $\langle L \rangle = 0$ (1-1),
two vortices enter at $\Omega \simeq 1.57$ (1-2), producing a
doubly-quantized vortex (1-3)~\cite{Lundh}.
Increasing the rotation frequency $\Omega$ further, we can nucleate two
more vortices in the condensate [(1-4) and (1-5)].
If we follow the branch starting from $\langle L \rangle = 2$ (2-1), the
two vortices escape out of the condensate with {\it increasing} $\Omega$
[(2-2) and (2-3)], and then four vortices enter at $\Omega \simeq 1.7$
(2-4).
It should be noted that the Bloch band picture holds even for the
nonrotating gas, while the ring-shaped profile manifests itself only if
the gas is rotating [cf. (1-1) and (2-3)].

The frequencies at which the Bragg reflection occurs in Fig.~\ref{f:2Dnoint}
(a) are different from those in Fig.~\ref{f:1D}.
The difference arises from the dispersion relation of the vortex states in
2D, i.e., the relation between the energy and angular momentum.
In the 1D case, the energy is given by $E_m^{\rm 1D} = \int [
\psi_m^{{\rm 1D}*} (-\partial_\theta^2 + i \Omega \partial_\theta)
\psi_m^{\rm 1D} + \gamma |\psi_m^{\rm 1D}|^4 / 2 ] d\theta =  m^2 - \Omega
m + \gamma / (4 \pi)$ for $\psi_m^{\rm 1D} = e^{i m \theta} /
\sqrt{2\pi}$.
The Bragg reflection between $\psi_m^{\rm 1D}$ and $\psi_{m + 2}^{\rm 1D}$
occurs when $E_m^{\rm 1D} = E_{m+2}^{\rm 1D}$, i.e., $\Omega = 2 m + 2$.
In the 2D case, we employ a variational wave function $\psi_m^{\rm 2D} =
(\pi |m|!)^{-1/2} d_m^{-|m| - 1} r^{|m|} \exp[-r^2 / (2 d_m^2) + i m
\theta]$ to obtain $E_m^{\rm 2D} = (|m| + 1) (d_m^{-2} + d_m^2) / 2 + K
(|m| + 1)(|m| + 2) d_m^4 / 4 - \Omega m + g (2 |m|)! / [2^{2 |m| + 2} \pi
|m|!^2 d_m^2]$, where the variational parameter $d_m$ is determined from
$\partial E_m^{\rm 2D} / \partial d_m = 0$.
When $K = 1$ and $g = 0$, the condition for Bragg reflections to occur
between $\psi_m^{\rm 2D}$ and $\psi_{m+2}^{\rm 2D}$ ($E_m^{\rm 2D} =
E_{m+2}^{\rm 2D}$) for $m = 0$, $2$, and $4$ is satisfied at $\Omega =
1.58, 1.83$, and $2.03$, respectively, which are in good agreement with
the numerical results in Fig.~\ref{f:2Dnoint} (a) (corresponding to the
points at which the red and blue curves cross with $\langle L \rangle$
changing by $\pm 2$).

The energy gap can also be obtained by a variational method.
The matrix element of the stirring potential, for example, between
$\psi_0^{\rm 2D}$ and $\psi_2^{\rm 2D}$ is calculated to be $|\int
\psi_2^{{\rm 2D}*} \psi_0^{\rm 2D} \varepsilon r^2 \cos 2\theta d{\bf r}|
\simeq 0.46 \varepsilon$ for $K = 1$.
The energy gap at $\Omega \simeq 1.57$ is then given by $\simeq 0.92
\varepsilon$, which also agrees very well with the numerical result
$\simeq 0.91 \varepsilon$.

The Bloch band structure in Fig.~\ref{f:2Dnoint} (a) indicates that if we
prepare the nonvortex state at $\Omega \lesssim 1.55$, and adiabatically
increase $\Omega$ to above $\Omega \simeq 1.6$, we obtain a
doubly-quantized vortex state by following the red curve.
Interestingly, we can nucleate vortices also by {\it decreasing} $\Omega$,
that is, by preparing a nonvortex state for $1.6 \lesssim \Omega \lesssim
1.7$ and decreasing $\Omega$ to below $\Omega \simeq 1.55$ by following
the second Bloch band (blue curve).
Figure~\ref{f:2Dnoint} (b) illustrates these adiabatic processes.
In order to examine the effect of dissipation on the adiabatic processes,
we replace $i$ with $i + \lambda$ on the left-hand side of
Eq.~(\ref{GP2D}), where a constant $\lambda$ is taken to be
$0.03$~\cite{Choi}.
We prepare the nonvortex ground state with $g = 0$, and gradually ramp up
$\Omega$ from $1.52$ to $1.6$ [red curve in Fig.~\ref{f:2Dnoint} (b)], or
ramp it down from $1.62$ to $1.54$ (blue curve).
As expected, two density holes with phase singularities come from
infinity, and unite to form a doubly-quantized vortex.
We note that the behaviors in Fig.~\ref{f:2Dnoint} (b) are almost the same
as for the dissipation-free case ($\lambda = 0$), indicating that the
process is robust against dissipation.
The energy gap between the first and second Bloch bands at $\Omega \simeq
1.57$ is $\simeq 0.036$ for $\varepsilon = 0.04$.
According to the Landau-Zener formula, the transition probability between
the energy gap, i.e., the probability that the nonvortex state remains so,
is below 1\% in the situation in Fig.~\ref{f:2Dnoint} (b), in agreement
with our numerical result.

\begin{figure}[tb]
\includegraphics[width=8.4cm]{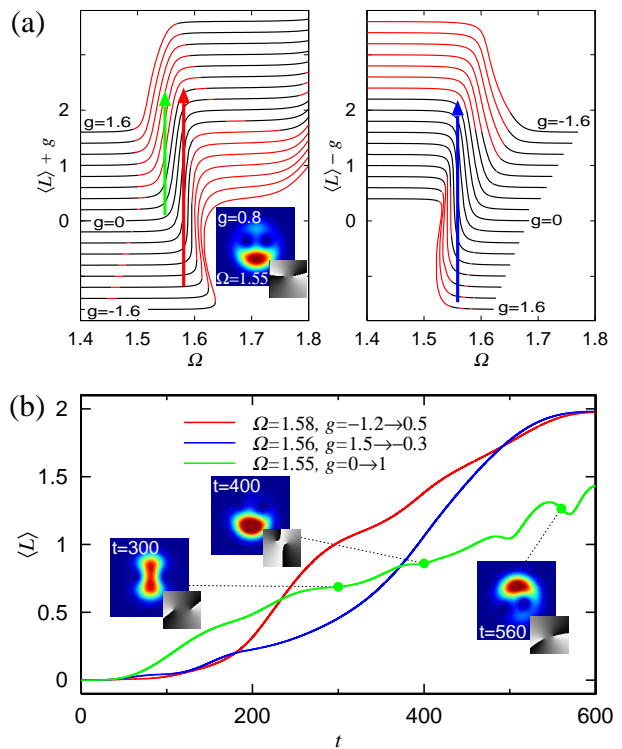}
\caption{
(a) The $g$ dependence of the first and second bands (left and right
panels) with $K = 1$ and $\varepsilon = 0.02$, where the curves are offset
by $\pm g$ for clarity.
The dynamically unstable regions are indicated by the red curves.
The inset shows a localized stationary state --``gap soliton''-- formed
near the first band at $g = 0.8$ and $\Omega = 1.55$.
(b) Time evolutions with $\lambda = 0.03$ and $K = 1$, where $g$ is
linearly ramped from $-1.2$ to $0.5$ with $\Omega = 1.58$ [red curve,
corresponding to the red arrow in (a)], from $1.5$ to $-0.3$ with $\Omega
= 1.56$ (blue curve, blue arrow), and from $0$ to $1$ with $\Omega = 1.55$
(green curve, green arrow).
A small perturbation to break the axisymmetry is added to the initial
state.
The change of $\varepsilon$ in the red and blue curves is the same as in
Fig.~\ref{f:2Dnoint} (b) , and $\varepsilon = 0.04$ is kept during $200
\leq t \leq 600$ in the green curve.
}
\label{f:2Dint}
\end{figure}

In the case of $g \neq 0$, the interaction bends the Bloch bands as shown
in Fig.~\ref{f:2Dint} (a), yielding hysteresis as in the 1D case.
We note that the region of $\Omega$ in which the Bragg reflection occurs
is shifted due to the interaction.
The shift is attributed to the difference in interaction energies
between the nonvortex and vortex states.
To show this we again employ the variational wave function $\psi_m^{\rm
2D}$.
We find that the condition for the Bragg reflection to occur between
$\psi_0^{\rm 2D}$ and $\psi_2^{\rm 2D}$, i.e., $E_0^{\rm 2D} = E_2^{\rm
2D}$, is satisfied at $\Omega = 1.53$ for $g = 1.6$ and $\Omega = 1.64$
for $g = -1.6$, in qualitative agreement with the shifts shown in
Fig.~\ref{f:2Dint} (a).
Since the interaction term of $E_m^{\rm 1D}$ is independent of $m$, the
shift does not occur in 1D (see Fig.~\ref{f:1D}).

The shift in the position of the Bragg reflection implies the possibility
of the adiabatic nucleation of vortices by changing the strength of
interaction using, e.g., the Feshbach resonance~\cite{Inouye}, at fixed
$\Omega$.
For example, at $\Omega \simeq 1.58$, the first Bloch band has angular
momentum $\langle L \rangle \simeq 0$ for $g \lesssim -1$ [left panel in
Fig.~\ref{f:2Dint} (a)], which continuously increases to $\langle L
\rangle \simeq 2$ with an increase in the interaction to repulsive (red
arrow).
Similarly, there is a region in which $\langle L \rangle$ changes from 0
to 2 as $g$ decreases from positive to negative [blue arrow in the right
panel in Fig.~\ref{f:2Dint} (a)].
No dynamical instability is present on the red and blue arrows.
Figure~\ref{f:2Dint} (b) illustrates these results, where the initial
state is the nonvortex ground state, and the strength of interaction is
gradually changed with $\Omega$ held fixed (red and blue curves).
We find that the two vortices are nucleated in a manner similar to that in
Fig.~\ref{f:2Dnoint} (b).
The repulsive-to-attractive case is particularly interesting because the
attractive interaction is usually considered to hinder vortex
nucleation~\cite{Lundh04}.

The adiabatic theorem breaks down when the path enters the region of
dynamical instability.
A notable example is shown as the green arrow in Fig.~\ref{f:2Dint} (a).
For $g \gtrsim 0.6$ at $\Omega \simeq 1.55$, a twofold symmetric state
[like the inset (1-2) of Fig.~\ref{f:2Dnoint} (a)] becomes dynamically
unstable against localization into the state shown in the inset of
Fig.~\ref{f:2Dint} (a).
This localization is due to the interplay between repulsive interaction
and negative-mass dispersion around the edge of the Brillouin zone.
The localized state can therefore be regarded as an analog of the gap
soliton in a periodic system~\cite{Meystre}.
The green curve in Fig.~\ref{f:2Dint} (b) demonstrates the time evolution
corresponding to the path shown as the green arrow in Fig.~\ref{f:2Dint}
(a).
The twofold symmetry as shown by the inset at $t = 300$ is broken in the
course of vortex nucleation, giving rise to two localized states as shown
by the insets at $t = 400$ and $t = 560$.

Figure~\ref{f:2Dint} (a) shows that a doubly-quantized vortex state with
$\langle L \rangle \simeq 2$ is dynamically unstable (stable) in most of
the region with $g < 0$ ($g > 0$).
This instability also gives rise to a localized state that undergoes
center-of-mass rotation.
We find from the numerical analysis that a doubly-quantized vortex state
with $g < 0$ evolves into the localized state, followed by the oscillation
between these two states in a manner similar to the split-merge
oscillations reported in Ref.~\cite{Saito}.

In conclusion, we have shown that the Bloch band structure arises in a BEC
confined in a harmonic-plus-quartic potential with a rotating drive, which
enables us to nucleate vortices adiabatically.
The physical mechanism of the vortex nucleation is very similar to the
Bragg reflection at the edge of the Brillouin zone.
Interestingly, we can nucleate vortices not only by increasing the
stirring frequency but also by decreasing it, or by changing the strength
of interaction at a fixed stirring frequency, which may be called a
``Feshbach-induced Bragg reflection''.
The adiabatic processes are robust against dissipation due, e.g., to a
thermal cloud.
Spontaneous localization of the rotating cloud is predicted, which signals
the existence of a gap soliton.

This work was supported by the Special Coordination Funds for Promoting
Science and Technology, a 21st Century COE program at Tokyo Tech
``Nanometer-Scale Quantum Physics'', and a Grant-in-Aid for Scientific
Research (Grant No. 15340129) from the Ministry of Education, Science,
Sports, and Culture of Japan.


\begin{thebibliography}{}
\bibitem{note}
Here vortices are defined as localized density depletions that have phase
singularities.
The core size of a vortex is kept from diverging in the limit of $g = 0$
due to the presence of a trapping potential.
\end{thebibliography}
\end{document}